\documentclass[sigconf]{acmart}

\acmConference[Arxiv]{}
\usepackage{booktabs} % For formal tables
\usepackage{hyperref}
\usepackage{amsmath}
\usepackage{verbatim}
\usepackage{siunitx}
\usepackage[shortlabels]{enumitem}
\usepackage{amsmath}
\usepackage{fancybox}
\usepackage{url}
\usepackage{flushend}
\usepackage{graphicx}
\usepackage{setspace}
\usepackage{soul}
\usepackage{multirow}
\usepackage[roman]{parnotes}
\usepackage{color}
\usepackage{colortbl}
\usepackage{listings}
\usepackage{tabulary}
\usepackage{xcolor}
\usepackage{framed}
\usepackage{subcaption}
\usepackage{comment}
\usepackage[justification=centering]{caption}
\usepackage{listings}
\usepackage{xcolor}

\definecolor{codegreen}{rgb}{0,0.6,0}
\definecolor{codegray}{rgb}{0.5,0.5,0.5}
\definecolor{codepurple}{rgb}{0.58,0,0.82}
\definecolor{backcolour}{rgb}{0,0,0}

\lstdefinestyle{mystyle}{
    backgroundcolor=\color{white},   
    commentstyle=\color{codegreen},
    keywordstyle=\color{codegreen},
    numberstyle=\tiny\color{codegray},
    stringstyle=\color{codepurple},
    basicstyle=\ttfamily\footnotesize,
    breakatwhitespace=false,         
    breaklines=False,                 
    keepspaces=true,                 
    numbers=none,                    
    numbersep=5pt,                  
    showspaces=false,                
    showstringspaces=false,
    showtabs=false,                  
    tabsize=2,
    captionpos=t
    frame=single
}

\definecolor{shadecolor}{RGB}{180,180,180}
\definecolor{codegreen}{rgb}{0,0.6,0}
\definecolor{codegray}{rgb}{0.5,0.5,0.5}
\definecolor{codepurple}{rgb}{0.58,0,0.82}
\definecolor{backcolour}{rgb}{0.95,0.95,0.92}
\definecolor{darkred}{rgb}{0.55, 0.0, 0.0}
\definecolor{darkpastelgreen}{rgb}{0.01, 0.75, 0.24}
\definecolor{darkmidnightblue}{rgb}{0.0, 0.2, 0.4}
\definecolor{jiri}{rgb}{0,0,255}
\definecolor{susie}{rgb}{150,0,150}
\definecolor{iftekhar}{rgb}{1, 0, 0}
\usepackage{pifont}% 

\makeatletter
\newcommand*{\rom}[1]{\expandafter\@slowromancap\romannumeral #1@}
\makeatother

\newcounter{observation}
\newcommand{\observation}[1]{\refstepcounter{observation}
	\begin{center}
		\Ovalbox{
			\begin{minipage}{0.93\columnwidth}
				{\bf  Observation \arabic{observation}:} #1
			\end{minipage}
		}
	\end{center}
}

\def\boldification #1 {\ifdraft\textbf{**#1**\newline\indent}\else\relax\fi}
\def\c #1 {}

\def\multilinerow #1 {\begin{tabular}[c]{@{}l@{}}#1\end{tabular}}

\definecolor{lightgray}{gray}{0.9}

\let\origthelstnumber\thelstnumber
\makeatletter
\newcommand*\Suppressnumber{%
  \lst@AddToHook{OnNewLine}{%
    \let\thelstnumber\relax%
     \advance\c@lstnumber-\@ne\relax%
    }%
}

\newcommand*\Reactivatenumber[1]{%
  \lst@AddToHook{OnNewLine}{%
   \let\thelstnumber\origthelstnumber%
   \setcounter{lstnumber}{\numexpr#1-1\relax}%
  }%
}

\setlist{nosep}
\setlist[itemize]{leftmargin=1em}
\setlist[enumerate]{leftmargin=3.3em}
\title{{Code Smells in Machine Learning Systems}}

\begin{document}

\author{Jiri Gesi}
\email{fjiriges@uci.edu}
\affiliation{%
  \institution{University of California, Irvine}
  \city{Irvine}
  \state{California}
  \country{USA}
 }
 
 \author{Siqi Liu}
\email{sliu17@uci.edu}
\affiliation{%
  \institution{University of California, Irvine}
  \city{Irvine}
  \state{California}
  \country{USA}
 }
 
 \author{Jiawei Li}
\email{jiawl28@uci.edu}
\affiliation{%
  \institution{University of California, Irvine}
  \city{Irvine}
  \state{California}
  \country{USA}
 }
 
 \author{Iftekhar Ahmed}
\email{iftekha@uci.edu}
\affiliation{%
  \institution{University of California, Irvine}
  \city{Irvine}
  \state{California}
  \country{USA}
 }
 \author{Nachiappan Nagappan}
\email{nnagappan@acm.org}
\affiliation{%
  \institution{Facebook}
  \city{Seattle}
  \state{WA}
  \country{USA}
 }

 \author{David Lo}
\email{davidlo@smu.edu.sg}
\affiliation{%
  \institution{Singapore Management University}
  \country{Singapore}
 }
 
  \author{Eduardo Santana de Almeida}
\email{esa@rise.com.br}
\affiliation{%
  \institution{Federal University of Bahia}
  \city{Salvador}
  \country{Brazil}
 }
 
 \author{Pavneet Singh Kochhar}
\email{pavneet.kochhar@microsoft.com}
\affiliation{%
  \institution{Microsoft Research}
  \city{Vancouver}
  \country{Canada}
 }
 
 \author{Lingfeng Bao}
\email{fjiriges@uci.edu}
\affiliation{%
  \institution{Zhejiang University}
  \state{Zhejiang}
  \country{China}
 }
 
% \maketitle
% \pagestyle{plain}

\begin{abstract}

As Deep learning (DL) systems continuously evolve and grow, assuring their quality becomes an important yet challenging task. Compared to non-DL systems, DL systems have more complex team compositions and heavier data dependency. These inherent characteristics would potentially cause DL systems to be more vulnerable to bugs and, in the long run, to maintenance issues. Code smells are empirically tested as efficient indicators of non-DL systems. Therefore, we took a step forward into identifying code smells, and understanding their impact on maintenance in this comprehensive study. This is the first study on investigating code smells in the context of DL software systems, which helps researchers and practitioners to get a first look at what kind of maintenance modification made and what code smells developers have been dealing with. Our paper has three major contributions. First, we comprehensively investigated the maintenance modifications that have been made by DL developers via studying the evolution of DL systems, and we identified nine frequently occurred maintenance-related modification categories in DL systems. Second, we summarized five code smells in DL systems. Third, we validated the prevalence, and the impact of our newly identified code smells through a mixture of qualitative and quantitative analysis. We found that our newly identified code smells are prevalent and impactful on the maintenance of DL systems from the developer's perspective.

\end{abstract}


\maketitle
\thispagestyle{plain}
\pagestyle{plain}

%!TEX root = main.tex

\section{Introduction}
%!TEX root = main.tex
\setcounter{page}{1}
% \boldification{{DL is important}}
In the past few years, Deep Learning (DL) systems, a branch of machine learning (ML), has become an inseparable part of billions of peoples’ lives worldwide, from personal banking to communication, from entertainment to transportation, and more~\cite{nlpExapmles,vc4auto}. Due to such ever-increasing dependence, ensuring DL system quality is of utmost importance. Failure to do so has already resulted in catastrophic consequences~\cite{teslaCrash}. 

As DL systems evolve and grow in size and complexity, continuous maintenance in the form of performance improvement, mandatory upgrades, and fixing bugs is necessary to ensure its correctness and continuous availability during its lifetime~\cite{sommerville1989approach}. However, maintenance of DL systems, similar to non-ML systems, can be hindered due to poor design and implementation choices. Compared to non-ML systems, DL systems are even more affected by maintenance issues since they are prone to the maintenance issues pertaining to both non-ML components and DL components as DL systems are combinations of both non-ML and DL components~\cite{sculley2015hidden}.

% \boldification{{many people have talked about any things CS is one of them which is useful for non-ML }}

Over the years, researchers have investigated the indicators of maintenance issues and methods to identify and quantify their impact~\cite{ahmed2017empirical,fowler1997refactoring,jebnounscent,sculley2015hidden}. Code smells is one such indicator, which is related to long term maintenance issues~\cite{fowler1997refactoring,jebnounscent,sculley2015hidden}. Prior research have investigated when and why code smells are introduced~\cite{tufano2015and} and how they evolve over time~\cite{arcoverde2011understanding,chatzigeorgiou2010investigating,lozano2007assessing,rapu2004using,tufano2015and}. Code smells' impact on software comprehensibility~\cite{abbes2011empirical}, fault-proneness, change-proneness~\cite{d2010impact, khomh2009exploratory, khomh2012exploratory}, and code maintainability~\cite{deligiannis2004controlled,li2007empirical,sjoberg2012quantifying,yamashita2012code, yamashita2013exploring} has also been demonstrated.

% \boldification{For DL some people have talked about CS but we don't still know about ????}

However, the majority of these studies focus on non-ML code smells with only a few focusing on ML code smells~\cite{jebnounscent} and none focuses on DL-specific code smells. Since DL and traditional software development is significantly different in terms of workflow and engineering practices~\cite{de2019understanding}, as well as DL's data dependent behavior~\cite{amershi2019software,wan2019does}, it is safe to assume that along with previously known code smells, there are code smells that are unique to DL systems which have not yet identified. 

\begin{figure*}[h]
    \centering
    \includegraphics[width=0.8\textwidth]{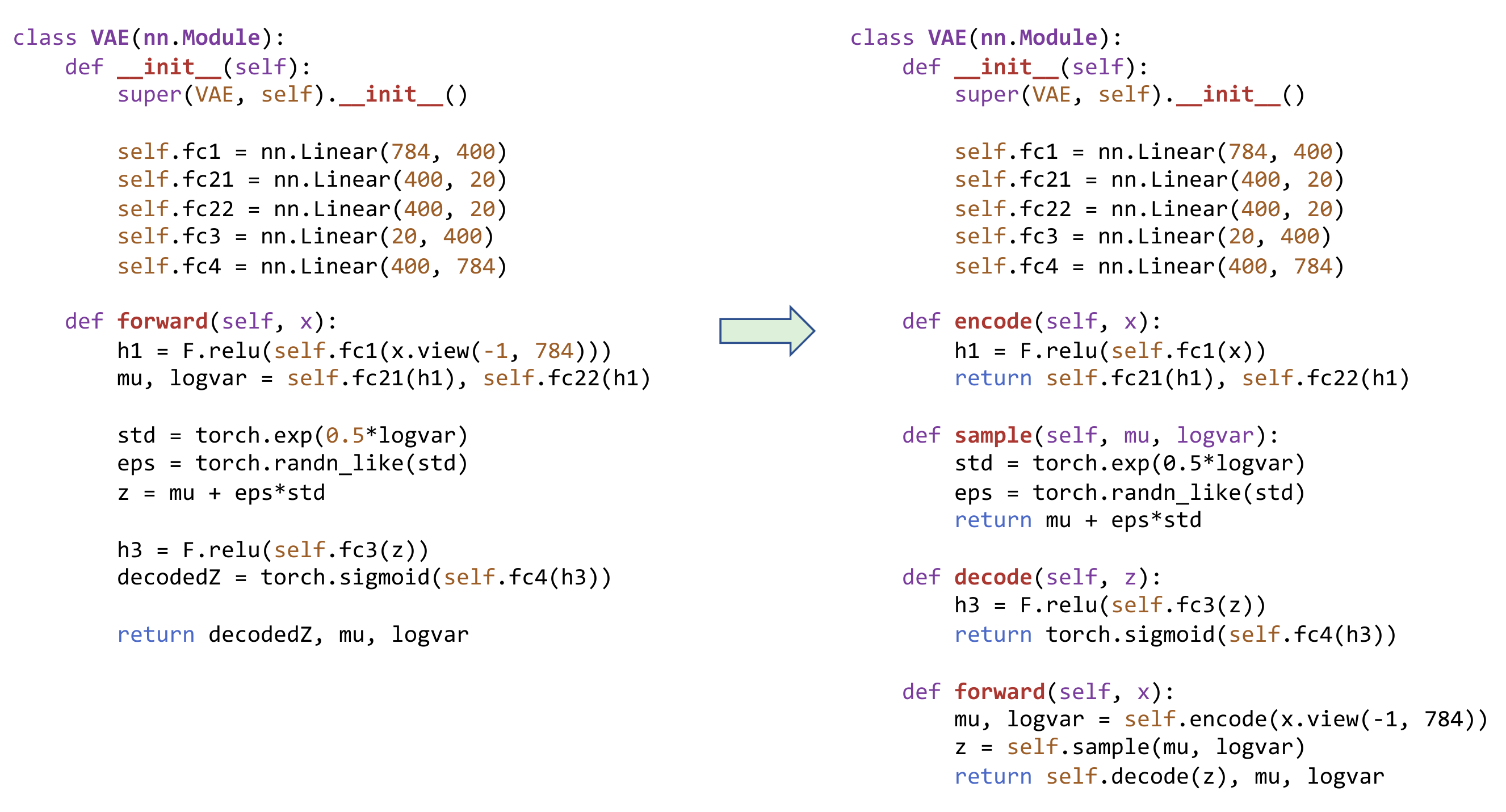}
    \caption{Jumbled Model Architecture code smell refactoring for Variational Autoencoeder model}
    \label{fig:code_sample}
\end{figure*}

A study conducted by Hadhemi et al.~\cite{jebnounscent} is closest to our work, where they studied code smells in DL systems. However, they investigated the prevalence of Python code smells; and analyzed code smells that were designed for non-DL general-purpose source code~\cite{chen2016detecting}. We posit that generic Python code smells provide only a partial picture, and there are DL-specific code smells that require further investigation. For example, Fig.\ref{fig:code_sample} shows an example of Jumbled Model Architecture (JMA) code smell where a Variational Autoencoder (VAE)~\cite{kingma2013auto} is extracted into encoding, sampling, and decoding\footnote{This refactoring commit is collected from the "NiftyNet"~\cite{NiftyNet} open-source software project.}. Intuitively, jumbled VAE impedes the understandability of model architecture and makes future maintenance difficult, this refactoring helps to ameliorate that. Due to the already proven impact of code smells on various aspects of non-DL software, it is safe to assume that code smells will have a similar, if not more detrimental effect on the long-term maintainability and overall quality of DL systems. Making it of utmost importance to get a complete picture of the unique code smells in DL systems and understand their impact. The first step towards achieving this goal is by identifying DL-specific code smells derived from real-world modifications applied to DL projects by developers.

% \boldification{to fill this gap, We have done ... }
In this study, we identify and analyze maintenance related modifications done by developers on 59 open source DL projects that were previously investigated by other researchers~\cite{jebnounscent}.

By employing a combination of PythonChangeMiner~\cite{pythonChangeMiner}, GitcProc~\cite{casalnuovo2017gitcproc} and manual analysis, we collected 426 maintenance related code changes from these 59 projects, where each change has at least three other similar occurrences among the projects. Next, using qualitative analysis, multiple coders independently coded collected changes into nine groups and extracted five frequently occurring code smells. Next, we validated the prevalence and severity of code smells by conducting a survey of 235 OSS DL developers. The survey analysis results show that our identified new code smells are often seen and have a significant impact on system maintenance activities.

In this paper, we answer the following research questions:

\noindent\textbf{RQ1}: What kinds of modifications do developers make frequently in DL systems?

\noindent\textbf{RQ2}: How prevalent are code smells in DL systems?

\noindent\textbf{RQ3}: How do practitioners perceive the identified code smells in DL systems?

The remainder of the paper is structured as follows. Section~\ref{sec:related} provides an overview of related work. Section~\ref{sec:method} details our methodology, with Section~\ref{sec:result} presenting our findings. Section~\ref{sec:discussion} places our results in the broader context of work to date and outlines the implications for DL practitioners and researchers. Section~\ref{sec:threats} lists the threats to the validity of our results. Section~\ref{sec:conclusion} concludes with a summary of the key findings and an outlook on our future work.

%!TEX root = main.tex
\section{Related Work}
\label{sec:related}

\begin{figure*}[h]
    \centering
    \includegraphics[width=0.9\textwidth]{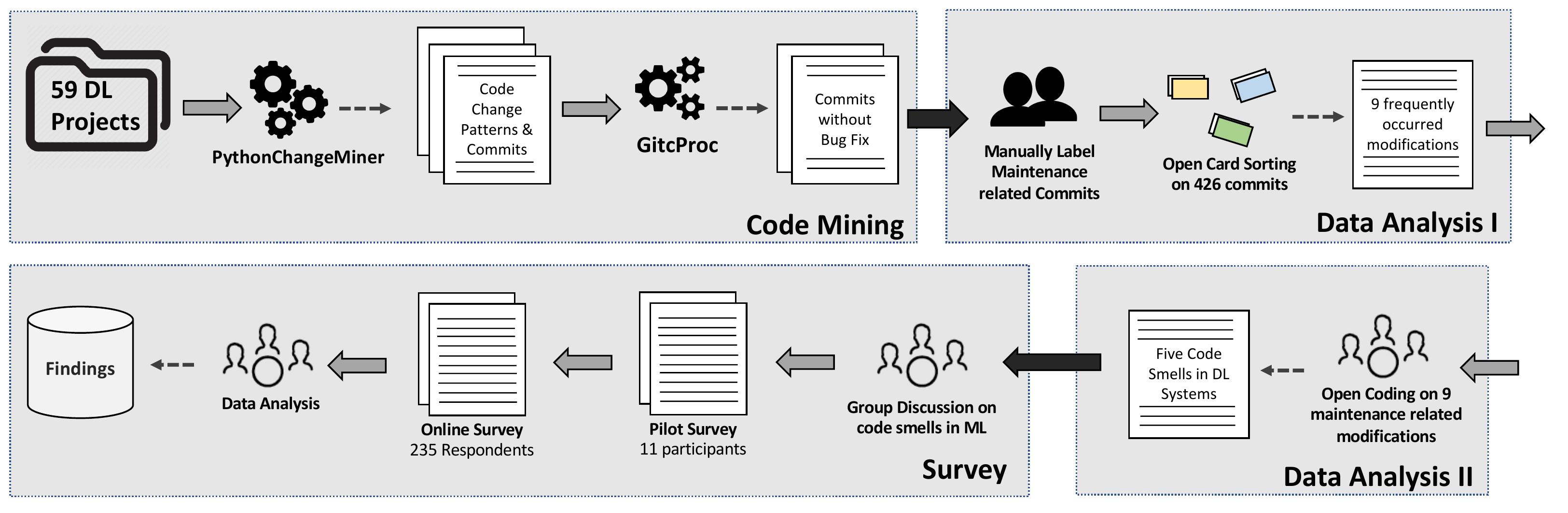}
    \caption{Schematic diagram of research methodology}
    \label{fig:methodology}
\end{figure*}

Code smells were introduced by Martin Fowler~\cite{fowler1997refactoring} to describe the design and implementation flaws in source code. These flaws do not make the software system behave incorrectly or crash but make it harder to understand, and maintain~\cite{d2010impact}. Research communities have investigated the impact of code smells in non-ML software systems such as how code smells impact fault-proneness and change-proneness~\cite{d2010impact, khomh2009exploratory, khomh2012exploratory}, it's impact on maintainability\cite{deligiannis2004controlled, li2007empirical, sjoberg2012quantifying, yamashita2012code,yamashita2013exploring}, when and why code smells are introduced~\cite{tufano2015and}, how they evolve over time~\cite{arcoverde2011understanding, chatzigeorgiou2010investigating, lozano2007assessing,rapu2004using, tufano2015and}, and how to detect code smells using different techniques~\cite{lanza2007object, moha2009decor,munro2005product, sahin2014code}. 

However, whether these code smells can capture all code smells relevant to DL systems is still an open question since existing research shows that there are significant differences between DL and traditional software systems. Wan et al. showed that the incorporation of DL into a software system significantly impacts the requirement analysis, system design, testing, and process management~\cite{wan2019does}. Scully et al. presented a set of unique anti-patterns in DL system development and highlighted a number of areas where technical debts unique to DL systems exist~\cite{sculley2015hidden}. Researchers also identified differences in the development process for DL systems due to the team formation and dependence on data which necessitates steps such as data understanding, data cleaning, model training, model deployment, and monitoring~\cite{amazonMLprocess, MSprocess, amershi2019software,de2019understanding}. All these differences can potentially introduce unique poor designs or implementations in source code, also known as code smells.

% The development of deep learning systems is a complex procedure and researchers have proposed various processes for DL development ~\cite{amershi2019software, amazonMLprocess,MSprocess}. These processes have some common steps: data analysis, data cleaning, building model, model deployment, and monitoring. During the data analysis stage, data scientists identify the areas of business that they can use 

%\boldification{We observe a lack of study about code smell in the context of the DL system. Prior Code smell-related work in ML systems, just one work done, but they used python smell on DL, which is not enough. But we need to see what smells they have done. After we identify new smells in the same repos, we can do a comparison in the discussion.}

Despite the clear differences between DL and traditional software systems, only a few studies have investigated code smells in the context of DL systems. Hadhemi et al.~\cite{jebnounscent} investigated the prevalence of Python code smells in DL systems along with investigating the differences in the distribution of code smells between DL and traditional systems. The code smells they investigated are: 

\textbf{Long Parameter List (LPL)}~\cite{fowler1997refactoring}: A method or a function that has a large number of parameters.

\textbf{Long Method (LM)}~\cite{fowler1997refactoring}: A method or a function that is extremely long.

\textbf{Long Scope Chaining (LSC)}~\cite{chen2018understanding}: A method or a function that has a deeply nested closure.

\textbf{Large Class (LC)}~\cite{brownantipatterns}: A class that has a large number of source code lines.

\textbf{Long Message Chain (LMC)}~\cite{brownantipatterns}: An expression for accessing an object using the dot operators through a long sequence of attributes or method calls.

\textbf{Long Base Class List (LBCL)}~\cite{brownantipatterns}: When a class extends too many base classes due to the multiple inheritances that Python language supports, it makes code hard to understand. 

\textbf{Long Lambda Function (LLF)}~\cite{brownantipatterns}: An anonymous function that is extremely long and complex in terms of conditions and parameters.

\textbf{Long Ternary Conditional Expression (LTCE)}~\cite{brownantipatterns}: A ternary conditional expression that is extremely long. 

\textbf{Complex Container Comprehension (CCC)}~\cite{brownantipatterns}: One-line comprehension list, set, or dictionary that contains a large number of clauses and filter expressions.

\textbf{Multiply-Nested Container (MNC)}~\cite{brownantipatterns}: a container (including set, list, tuple, dict) that is deeply nested.

%\boldification{This code smells only for the code part, but the DL system contains code, model, and data, so we need to identify code smells for the rest two parts in the deep learning system.}

As it can be seen for the definitions, these code smells were designed for traditional general-purpose Python code~\cite{chen2016detecting}. However, in a DL system, there is general-purpose code, along with model architecture, data preparation, and pipeline related code. Hence, we posit that there are other code smells that are unique to DL specific code (i.e., model architecture, data preparation and pipeline, etc.) 

Prior research in the context of non-ML systems indicated that developers have differentiated opinions about code smells, their prevalence, and effect~\cite{yamashita2012code}. However, existing research in DL did not investigate how developers perceive code smells in the context of DL systems. As result, questions such as how prevalent these smells are, and how developers perceive their impact remains unanswered. We aim to fill the gap and answer the questions in this work.

%!TEX root = main.tex
\section{Methodology}
\label{sec:method}

 \begin{figure*}
  \centering
    \includegraphics[width=1.0\textwidth]{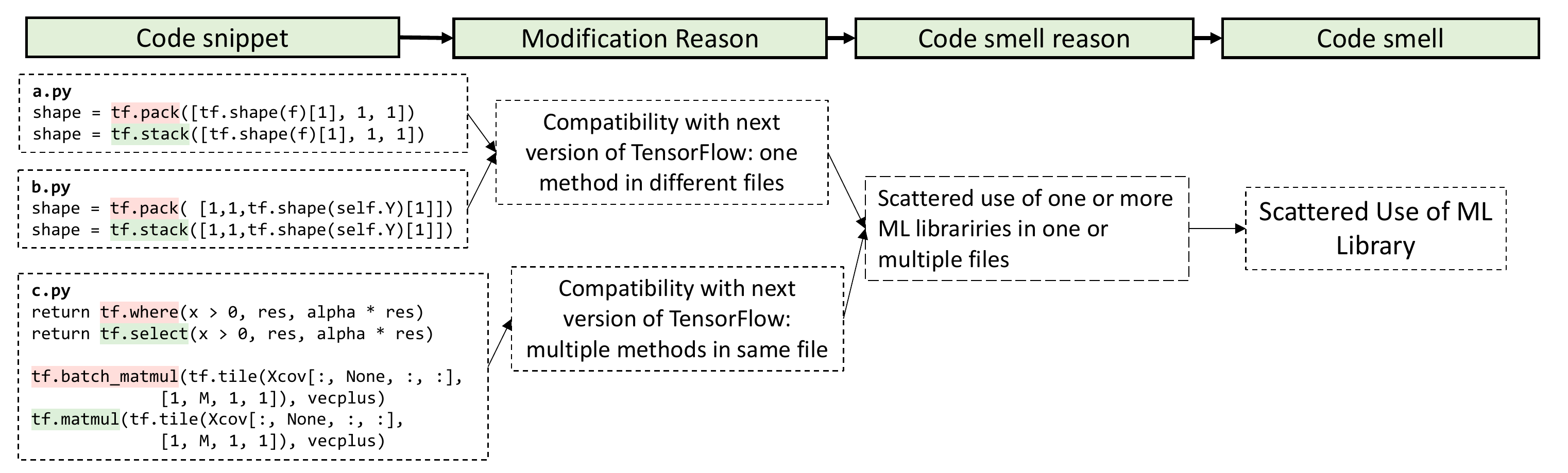}
  \caption{An example of using qualitative analysis for determining code smells in deep learning systems}
  \label{fig:example_determine2}
\end{figure*}

We used a mixed method approach consisting of mining software repositories and qualitative analysis. Figure~\ref{fig:methodology} shows the process that we follow in this study. We start by code mining to gather recurring code change patterns, then apply open card coding to identify new code smells, and finally, conduct a large-scale survey to validate the prevalence and impact of the newly identified code smells.

\subsection{Code Mining}
\label{sec:mining}

Our first step was collecting recurring code changes in 59 open source DL systems. These projects were investigated by Hadhemi et al.~\cite{jebnounscent} in their study and we wanted to investigate whether there are other codes smells unique to DL in these systems besides generic Python code smells, thus we used the same dataset. 
%The detailed steps show in the Code Mining part of Fig.~\ref{fig:methodology}.

\subsubsection{Data Collection}
\begin{figure}[h]
    \centering
    \begin{subfigure}{0.8\columnwidth}
        \includegraphics[height=0.27in]{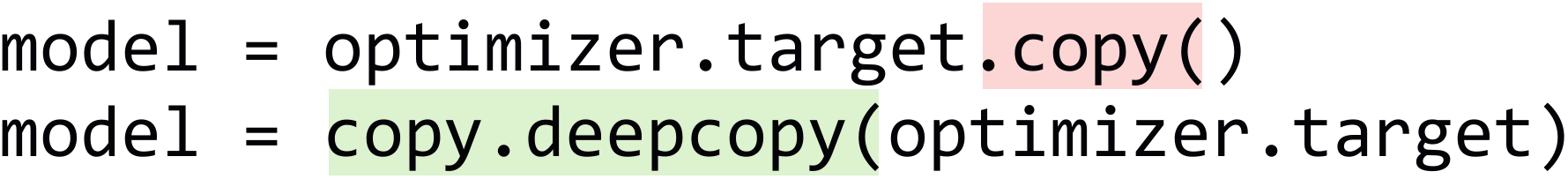}
        \caption{Code snippet 1}
        \label{subfig:correct}
        
    \end{subfigure}
    
    \begin{subfigure}{0.8\columnwidth}
        \includegraphics[height=0.27in]{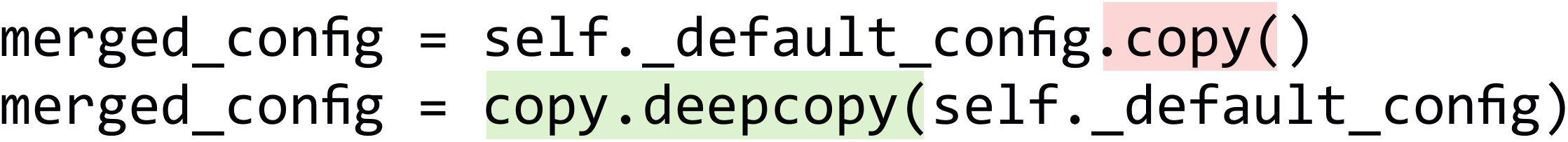}
        \caption{Code snippet 2}
        \label{subfig:notwhitelight}
    \end{subfigure}

    \begin{subfigure}{0.8\columnwidth}
        \includegraphics[height=0.27in]{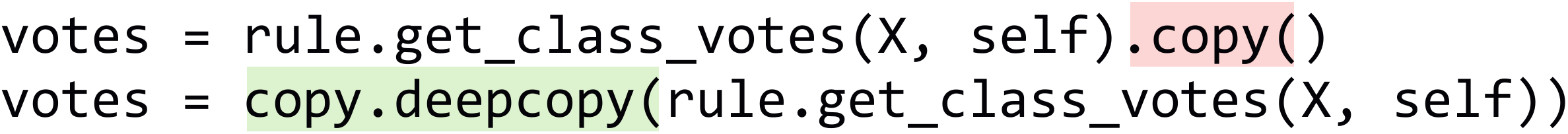}
        \caption{Code snippet 3}
        \label{subfig:nonkohler}
    \end{subfigure}

   \caption{An example of a changing pattern identified in several projects on GitHub. The developers switched from using built-in copying to creating a deep copy of an object using a \texttt{copy} module of a Standard library.}
   \label{fig:example_pattern}
\end{figure}

We started by obtaining $90,301$ commits from the 59 DL open source projects downloaded on May 20, 2020. Next, we used PythonChangeMiner~\cite{pythonChangeMiner} to detect and group commits with similar change patterns. PythonChangeMiner mines the history of a given repository using the PyDriller framework~\cite{PyDriller} and builds change graphs for matching functions in each changed file for a commit. To achieve this, both versions of the file (before and after the change) are parsed into Abstract Syntax Trees (ASTs)~\cite{nguyen2019graph}, which are then traversed to create the structure of \textit{fine-grained Program Dependence Graph (fgPDG)}. Then, the obtained fgPDG are analyzed to find all node pairs before and after the change using GumTree~\cite{lam2006gumtree}, resulting in grouped change pattern categories. Figure~\ref{fig:example_pattern} shows an example of a changing pattern identified in several projects that developers switched from using built-in copying to creating a deep copy of an object using a \textit{copy} module. To make sure that our analyzed patterns are common across multiple projects and not specific to a project, we extracted code changes that happened at least three times within all commits across multiple projects. We identified $1,942$ commits matching this criterion.

Research shows that refactoring (non-bug-fixing and non-program behavior-altering commits) is performed to remove code smells~\cite{herzig2013s}. Since our identified $1,942$ patterns contained both bug-fixing, non-bug-fixing commits, we removed bug-fixing commits from our analysis as they alter program behavior. We used GitCProc~\cite{casalnuovo2017gitcproc} for this purpose, which identifies the bug-fixing commits based on the presence of specific words in the commit message. Words such as \textit{error, bug, defect} and, \textit{fix} are considered while identifying bug-fix commits by GitCProc. After removing bug-fix related commits, $1,335$ non-bug-fixing commits were left, which come from all 59 projects. Next, the first and second authors independently went through the commits to identify the commits related to maintenance. They relied on the commit message and compared the code before and after the update for deciding whether the commit was maintenance related or not. They initially used 10\% (134) of the commits and independently labeled them. After initial labeling, the inter-rater agreement was 0.61, which according to Landis et al.\cite{landis_1975} is considered as a substantial level of agreement. After an initial disagreement on some of the commits, the authors discussed their approach and had a complete agreement regarding the label of commits initially disagreed. Then the two authors labeled the remaining 1,201 commits together. This resulted in selecting $426$ maintenance related commits where each commit had at least three occurrences across multiple projects.
\subsubsection{Modification Category Creation} 
Our next step was to group these commits based on the modification reasons. To do so, we followed descriptive coding~\cite{saldana2015coding} which is used for identifying topics from data. The result of descriptive coding is categorized groups based on identified topics. Two authors jointly conducted the descriptive coding on the selected 426 commits. They relied on the commit message and compared the code before and after the update for identifying the reasons for making the changes. This resulted in grouping the commits into nine modification categories. We selected descriptive coding technique for the following reasons: (1) we can get an overview of recurring changes that are indicative of poor maintainability; (2) we can obtain the context of these modifications.

\subsubsection{Code Smell Categorization} 

Our primary goal was to extract code smells from the frequently occurred modifications. For this purpose, in the next step first and the second authors checked if the modification reasons mentioned in Table~\ref{tab:modifications} met the following criterion: (1) whether the modification reason is general (common to many DL systems), (2) if there is a general solution to the root cause that required the modification. If both criteria are met, they considered the modification reason as a code smell. Figure~\ref{fig:example_determine2} shows an example of qualitative analysis to determine whether a modification is a code smell.

Two rounds of descriptive coding were conducted. In the first round, the first and second authors independently investigated all modification reasons and created a list of code smell candidates based on the previously mentioned two criteria. After discussing they curated a list of 12 code smells and reached an inter-rater agreement of $83.2\%$. In the second round, these 12 code smells were presented to all authors, and after discussion, everyone agreed on five new code smells and the remaining seven were discarded as they did not meet the previously mentioned criteria completely.

Since the collected commits consisted of both new code smells and pre-existing code smells that were identified by Hadhemi et al~\cite{jebnounscent} (listed in Section\ref{sec:related}), we check the prevalence of generic Python code smells among the $427$ commits. Since Pysmell~\cite{chen2016detecting} can identify these smells, instead of applying quantitative techniques, we relied on Pysmell for this purpose. We ran Pysmell before and after applying each of the $426$ commits and calculated the number of fixed generic Python code smells. If the count of code smells decreased, we labeled that commit as Python code smell fixing commit. Through this analysis, we identified eight Python code smells in our data.

%: Long Parameter List (LPL), Long Method (LM), Long Scope Chaining (LSC), Large Class (LC), Long Message Chain (LMC), Long Base Class List (LBCL), Long Lambda Function (LLF), Long Ternary Conditional Expression (LTCE), Complex Container Comprehension (CCC), and Multiply-Nested Container (MNC). 

\subsection{Survey}

We delivered a survey to gain an understanding of the prevalence and severity of the newly identified code smells and gather the developer's perspective about them. 

\subsubsection{Protocol} We based our questions on the identified code smells from code change pattern mining. Our questionnaire included questions about the following topics (the complete questionnaire is available as supplemental material\footnote{https://github.com/codesmell-material/codeSmell}):

\begin{itemize}
\setlength{\itemindent}{.1in}

\item \textit{Demographics}: We asked questions about organizations, geographical locations, and ML-related working experiences for this part of the survey. 

\item \textit{Self-perception}: We let respondents self-identify their professional categories ("I think of myself as a/an..." like researchers, engineers, scientists, etc). We used the answers to classify all respondents into four groups based on the result's keywords: data scientist/engineer, Machine Learning (ML) engineer, software engineer, and project manager based on their self-perception. ML engineers sit at the intersection of software engineering and data science, whose job is applying ML techniques and developing DL models. Data scientists/engineers are the group of people who create and maintain optimal data pipeline architecture, study and understand the data, and clean data. All respondents who are working on data-related jobs are grouped. Software engineers are those who build the software system and deploy the DL models.

\item \textit{Perception on code smells}: We asked respondents whether they have encountered the code smells. To clarify any possible confusion, we provided a definition and a simple example for each code smell. If they responded ``yes'', we also asked them to what extent the code smell impacts their DL system maintenance (Very Serious, Serious, Moderate, Scarcely, and Not At All). 
% \item \textit{Other challenges}: We asked an open-ended question to understand if there are any additional code smells that make the DL systems maintenance more difficult:``What source code modification challenges do you frequently face when maintaining the DL system?'' 
\end{itemize}

We followed a pilot protocol~\cite{brace2018questionnaire} while designing the survey. We designed a pilot version and sent them to a small subset of developers (11 developers). Based on the feedback, we rephrased some questions to make them easier to understand. We simplified and merged some questions to ensure that participants could finish the survey in 7 minutes. The responses from the pilot survey were used solely for improving the survey questions and were not included in the final results. We also translated our original survey to a Chinese version to support respondents who read Chinese before distributing the survey. two of the authors (one of them is a native English speaker and the other a native Chinese speaker) discussed the survey and performed the translation together.

\subsubsection{Respondent Selection} We aimed to get a sufficient number of practitioners from diverse backgrounds working on open source DL development and maintenance. Thus, we collected active contributors' emails in the 59 DL projects by using GitHub REST APIs. In total, we collected 1,157 email addresses and successfully distributed them to 1,061 contributors. We kept the survey anonymous, but the respondents could choose to receive a summary of the study.

\begin{figure}
  \centering
    \includegraphics[width=1.0\columnwidth]{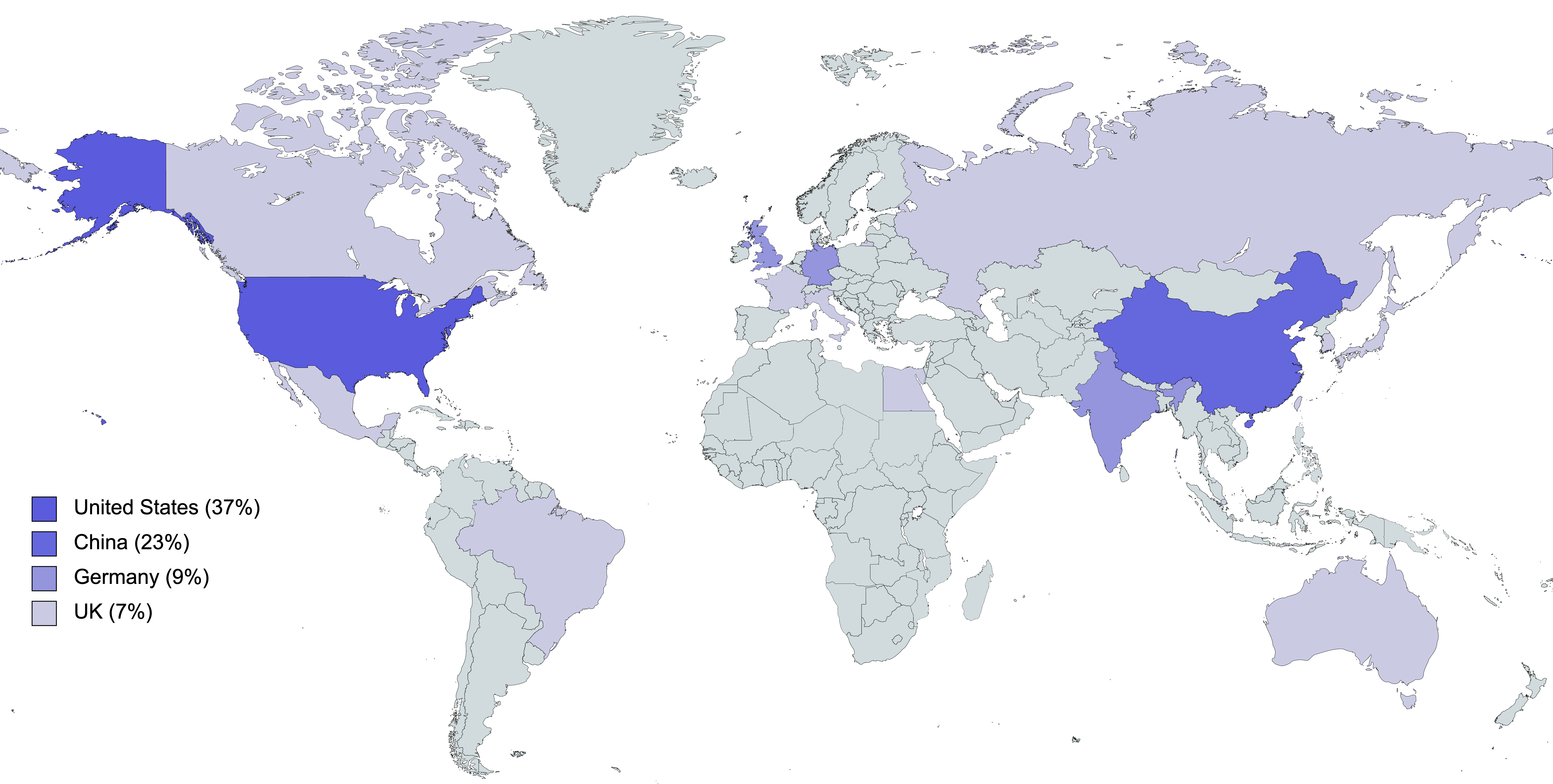}
  \caption{Countries in which survey respondents reside. The legend presents the top 4 countries with most respondents}
  \label{fig:surveyMap}
\end{figure}

In total, we received 265 responses. After excluding incomplete surveys, 235 responses were considered valid. The countries and the corresponding number of respondents are shown in Fig.~\ref{fig:surveyMap}. The survey respondents who met our criteria are distributed across 15 countries and six continents. The majority of our respondents currently work in North America, Asia, and Europe, with the United States and China being the top two countries. Respondents' software development experience varies from 1 to 23 years with an average of 5.25 years, and their DL development experience varies from 1 to 10 years with an average of 3.13 years.

\subsubsection{Survey Data Analysis}

To analyze the responses, we used descriptive statistics. 

For the 235 valid responses to the question related to whether they have encountered our identified code smells, we normalized the frequency of each code smell by computing the percentage of respondents who have encountered code smells. If a high proportion of respondents reported that they have encountered a certain code smell, we consider this smell as more common. We did the same for the impact level of the code smell question. We also analyzed the responses based on roles. We mainly analyzed the responses from the top three categories of respondents which belonged to software engineers, ML engineers and data scientist/engineers since the number of project manager respondents were too few.

To check if there is a significant difference between new identified code smells in terms of impact, we adopted the Scott-Knott test~\cite{jelihovschi2014scottknott}. Scott-Knott test divides the measurement averages into statically distinct groups by hierarchical clustering analysis. However, the limitations of the Scott-Knott test are that it assumes the data are in a normal distribution and it may create groups with trivially different from each other. Thus, we adopted its normality and effect size-aware variant Scott-Knott effect size difference (ESD) test~\cite{tantithamthavorn2016empirical}. The Scott–Knott ESD test (1) corrects the normal distribution of the input data and (2) merges any two statistically different groups of negligible effects. A detailed description of the Scott–Knott ESD test can be found in ~\cite{tantithamthavorn2016empirical}.

%!TEX root = main.tex

\section{Results}
\label{sec:result}

In this section, we report the answers to our targeted research questions and findings that emerged from the data.

\begin{table*}[hbt]
\centering
\caption{Summary of modification commits and their distribution}
\label{tab:modifications}
\resizebox{0.95\textwidth}{!}{%
\begin{tabular}{|l|l|c|}
\hline
\rowcolor[HTML]{FFFFFF} 
\textbf{Modification Categories} &
  \textbf{Modification reasons} &
  \textbf{\begin{tabular}[c]{@{}l@{}}Percentages  \\ of selected \\ commits (\%)\end{tabular}} \\ \hline
\rowcolor[HTML]{FFFFFF} 
\textit{Change function declaration} &
  \begin{tabular}[c]{@{}l@{}} Rename functions; Change lambda function to normal functions;\\ Change function signatures; Convert public function to private;\\ Convert private function to public.  \end{tabular} & $21\% $
   \\ \hline
\rowcolor[HTML]{FFFFFF} 
\textit{Update/replace ML library} &
  \begin{tabular}[c]{@{}l@{}} Update deprecated functions with new ML library; Resolve \\python version compatibility issue;  Resolve ML library compati-\\-bility issue; Switch to new ML library; Employ another ML \\library to improve model performance. \end{tabular} & $19\%$
   \\ \hline
\rowcolor[HTML]{FFFFFF}
\begin{tabular}[c]{@{}l@{}}\textit{Remove redundant }\\ \textit{debugging code}\end{tabular}&
  \begin{tabular}[c]{@{}l@{}} Clean up no longer used debugging of redundant code. \end{tabular} & 16\%
   \\ \hline
  \rowcolor[HTML]{FFFFFF} 
\textit{Replace hard-coded value}&
  \begin{tabular}[c]{@{}l@{}} Replace hard-coded model names, learning rate, parameters, num-\\-bers, and etc to variables. \end{tabular} & 13\%
   \\ \hline
\rowcolor[HTML]{FFFFFF} 
\textit{Extract class/function} &
  \begin{tabular}[c]{@{}l@{}} Create a separate new class/function to remove old duplicate \\code; Isolate independent parts of code; Split long function/class. \end{tabular} & 9\%
   \\ \hline
\rowcolor[HTML]{FFFFFF}  
\textit{Data preparation modification} &
  \begin{tabular}[c]{@{}l@{}} Resolve Data API compatibility issue; Separate data preparation \\code; Clean up data loader.   \end{tabular} & 8\%
   \\ \hline
\rowcolor[HTML]{FFFFFF} 
\textit{Move code} &
  \begin{tabular}[c]{@{}l@{}} Move code to proper files or positions; Simplify deep nested closure\\ functions or containers. \end{tabular} & 6\%
   \\ \hline
\rowcolor[HTML]{FFFFFF} 
\textit{Model architecture modification} &
  \begin{tabular}[c]{@{}l@{}} Rewrite model architecture source code; Deep learning layers\\and  parameter modification; Replace with a new model; Separate \\model parts.   \end{tabular} & 6\%
   \\ \hline
   \rowcolor[HTML]{FFFFFF} 
   
\textit{Remove dispensable dependency} &
  \begin{tabular}[c]{@{}l@{}} Remove unused or dispensable DL library/frameworks; Resolve\\ dependent conflicts. \end{tabular} & 2\%
   \\ \hline
   
\end{tabular}%
}
\end{table*}

\subsection{Maintenance Related Modifications in Deep Learning}
\textbf{RQ 1: What kinds of modifications do developers make frequently in DL systems?}

To answer this question, we mined 59 open source DL project repositories and identified 426 maintenance-related modification commits. By using descriptive coding, we categorized the selected commits into nine modification categories (explained in Section~\ref{sec:method}). The modification categories, identified modification reasons, and their corresponding distributions are shown in Table~\ref{tab:modifications}. 

Our manual analysis revealed that, as expected, some of the frequent modifications are specific to DL systems and others are not. For example, the most frequent (21\%) modification category named \textbf{Change function declaration} which involves renaming functions, changing lambda functions to normal functions, and modifying function signatures is not specific to DL. \textbf{Extract class/function} category which includes changes pertaining to separating new class, isolating independent parts of code, and splitting long functions is also not specific to DL.

We also found that three of the modification categories are specific to DL. 

%\subsubsection{DL specific modifications}

%We found following three modifications that are specific to DL systems: %There are repetitively occurring modifications that only exist in DL systems, such as \textit{update/replace ML libraries}, \textit{data preparation changes}, and \textit{model architecture changes}.

%\begin{itemize}
%\setlength{\itemindent}{.2in}

%\item 
\textbf{Update/replace ML library}: This recurring modification is the second most frequent category of modification (19\%). Similar to API update/replace in traditional systems, developers usually use third-party DL libraries and frameworks to implement DL functionalities. However, DL libraries are usually updated more frequently than traditional libraries~\cite{sculley2015hidden} and DL developers need to fix either deprecated or outdated functions to keep up with the updates. For example, the code snippet in Figure~\ref{fig:example_determine2}, shows that developers had to replace API names to resolve the compatibility issue with a newer version of TensorFlow.

\textbf{Data preparation modification}: This recurring modification is performed on the data preprocessing steps. We found that 8\% of overall modifications in our dataset belonged to this group. Since a substantial part of code in DL systems is written for data preparation, and feeding to DL model any changes to the data source, preparation steps, or the model architecture requires this category of modification. 

% since the data is highly specific to the project context, 
% As DL model architecture or data source changes, the data preprocessing method has to change accordingly. 

%\item 
\textbf{Model architecture modification}: This recurring modification is done on DL model architecture related code. In order to resolve model degradation problems, developers iteratively train models or deploy new model architectures. We also found that developers make modifications to improve the model architecture by untangling the components. This group of modifications is 6\% of our analyzed commits.

%\end{itemize}

%\observation{\textit{Update/replace ML libraries} is the most frequently Some modification types are specific to DL systems, such as , \textit{Data preparation modification}, and \textit{Model architecture modification}.}

\observation{One third (33\%) of the maintenance related modifications in DL systems are specific to DL systems and are related to the data, model, and library.}

Interestingly, our results highlight another category of  modification that is not specific to DL, but contains some DL specific changing reasons:

\textbf{Replace hard-coded value}: This is the recurring modification where developers replace hard-coded values with variables. Similar to traditional software, hard-coded values make it difficult to maintain software systems. We found developers frequently replace hard-coded model path, hyper-parameters, and learning rate with variables. 13\% of our identified commits fall into this category.

\textbf{Remove redundant debugging code}: Developers frequently remove unnecessary debugging code in DL systems. The software engineering community has developed a number of tools, IDEs, and techniques to help catch bugs. Unfortunately, practitioners for DL systems do not enjoy the same robust set of debugging tools available for traditional software while debugging DL models due to the opaqueness of DL models and strong coupling between model and software components~\cite{sculley2015hidden}. Thus, many DL developers resort to using print statements for debugging. 16\% of maintenance-related modifications were grouped into this category.

%The redundant debugging code can negatively impact the program's understandability. 

\textbf{Move code}: In this category of recurring modification, developers move code between files and positions. Developers often put model training, testing, and validation related code in the same file. Later on, they end up moving each of the training, testing, and validation to separate files. We found that 6\% of the modification commits belong to this category.

\textbf{Remove dispensable dependency}: This is the recurring modification where developers remove unused or unnecessary dependencies. Resolving dependency compatibility problems or versioning conflicts can be time consuming. As a result, developers are usually reluctant to remove dispensable dependencies until they have to. This kind of modification consists of 2\% of modifications commits in our dataset.

%\observation{\textit{Replace hard-coded value, Remove unwanted debugging code, Move code to proper places}, and \textit{Remove dispensable dependency} modifications have some specific changing reasons to DL systems.}

\begin{table*}[hbt]
\centering
\caption{Summary of Newly Identified Code Smells in Deep Learning Systems}
%(total number of selected commits is 426 and number of projects is 59)}
\label{tab:new_smells}
\resizebox{\textwidth}{!}{%
\begin{tabular}{|
>{\columncolor[HTML]{FFFFFF}}l |
>{\columncolor[HTML]{FFFFFF}}l |
>{\columncolor[HTML]{FFFFFF}}c |
>{\columncolor[HTML]{FFFFFF}}c |}
\hline
{\color[HTML]{000000} \textbf{Code Smell}} &
  {\color[HTML]{000000} \textbf{Signs and symptoms}} &
  {\color[HTML]{000000} \textbf{\begin{tabular}[c]{@{}c@{}}Percentages\\ of\\ commits (\%)\end{tabular}}} &
  {\color[HTML]{000000} \textbf{\begin{tabular}[c]{@{}c@{}}Percentages of \\ Projects (\%)\end{tabular}}} \\ \hline
{\color[HTML]{000000} \textit{\begin{tabular}[c]{@{}l@{}}Scattered Use of ML \\ Library\end{tabular}}} &
  {\color[HTML]{000000} \begin{tabular}[c]{@{}l@{}}Scattered use of ML API in multiple files, once an ML API needs \\to be modified/updated, and the practitioners must modify places \\across several files.\end{tabular}} &
  {\color[HTML]{000000} 13\%} &
  {\color[HTML]{000000} 54\%} \\ \hline
{\color[HTML]{000000} \textit{\begin{tabular}[c]{@{}l@{}}Unwanted Debugging\\ Code\end{tabular}}} &
  {\color[HTML]{000000} \begin{tabular}[c]{@{}l@{}}A debugging code fragment, method, or class is no longer used, \\ but  still is left in the source code\end{tabular}} &
  {\color[HTML]{000000} 17\%} &
  {\color[HTML]{000000} 41\%} \\ \hline
{\color[HTML]{000000} \textit{Deep God File}} &
  {\color[HTML]{000000} \begin{tabular}[c]{@{}l@{}}A file contains multiple components of a DL system, such as mo-\\-del training, testing, etc\end{tabular}} &
  {\color[HTML]{000000} 9\%} &
  {\color[HTML]{000000} 37\%} \\ \hline
{\color[HTML]{000000} \textit{\begin{tabular}[c]{@{}l@{}}Jumbled Model\\ Architecture\end{tabular}}} &
  {\color[HTML]{000000} \begin{tabular}[c]{@{}l@{}}DL model architecture parts are cobbled together and difficult to \\understand and maintain.\end{tabular}} &
  {\color[HTML]{000000} 6\%} &
  {\color[HTML]{000000} 19\%} \\ \hline
{\color[HTML]{000000} \textit{\begin{tabular}[c]{@{}l@{}}Dispensable \\ Dependency\end{tabular}}} &
  {\color[HTML]{000000} \begin{tabular}[c]{@{}l@{}}An installed DL library or framework is no longer used or can be\\ replaced by other existing ones\end{tabular}} &
  {\color[HTML]{000000} 2\%} &
  {\color[HTML]{000000} 9\%} \\ \hline
\end{tabular}%
}
\end{table*}

\subsection{Code smells in Deep Learning System}

\textbf{RQ 2: How prevalent are code smells in DL systems?}

Through manual analysis of the maintenance related modifications done on real-world projects, we identified five code smells in DL systems (details in Section \ref{sec:method}). Table~\ref{tab:new_smells} shows the five code smells along with their signs and symptoms ordered based on the frequency of occurrence in projects from high to low.

\begin{figure}[h]
  \centering
    \includegraphics[width=0.9\columnwidth]{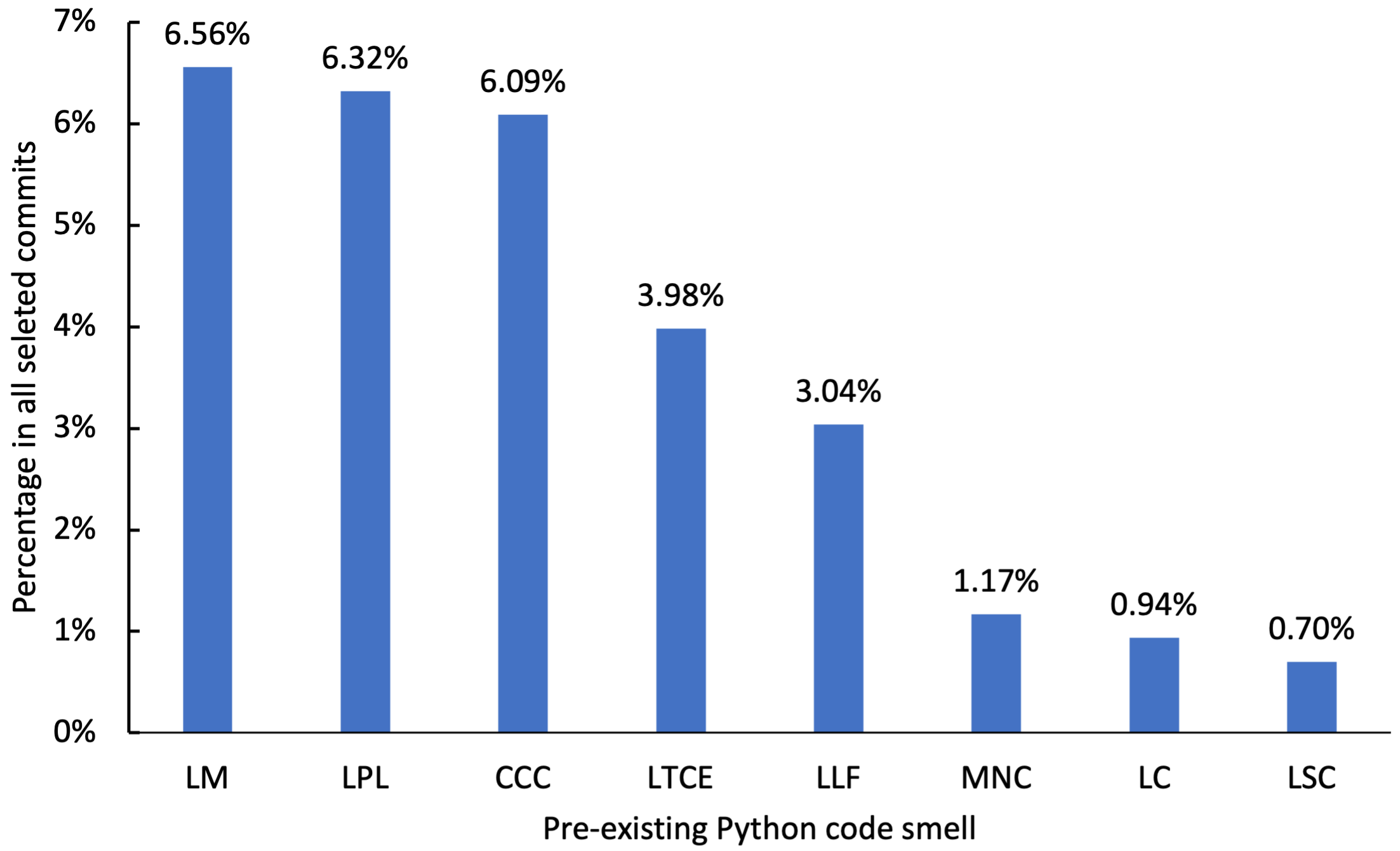}
  \caption{Prevalence of identified existing code smells }
  \label{fig:exsiting_smells}
\end{figure}

\textbf{Scattered Use of ML Library}: This smell is about implementing third-party ML libraries/frameworks in a non-cohesive manner throughout the project. As a result, whenever these libraries/frameworks update, developers have to modify multiple positions in single or multiple files. Such scattered use of ML library requires additional effort from the developer while maintaining the source code. 32 out of 59 (54\%) projects have at least one commit showing this problem. 

\textbf{Unwanted Debugging Code}: This smell was derived from the recurring pattern of leaving unwanted or unnecessary code in the DL system and we found 24 out of 59 DL projects have this code smell. DL systems tend to be more complicated than a traditional system and developers use debugging code for getting data shape or printing current status to understand the code. However, left uncleaned these debugging codes can impede maintainability. If many people are working on a project, individuals are more reluctant to remove code that they do not thoroughly understand since no one wants to be responsible for errors. With these redundant codes left in the system, the code will be more difficult to understand, especially for DL systems.

\textbf{Deep God File}: This smell was derived from the recurring pattern where developers kept separating DL parts into multiple files after they had initially put some or all of them into one big file. We found 22 projects (37\%) with this code smell. \textit{Deep God File} usually starts small, but over time, they get bloated as practitioners may find it mentally easier to place programs into existing files.

\textbf{Jumbled Model Architecture}: This smell was derived from the recurring pattern when DL practitioners programmed the DL models, they do not clearly divide the different functional parts of the model. Instead, all parts of the model are jumbled together, which makes model code difficult to understand. We found 11 (19\%) projects with this code smell.

\textbf{Dispensable Dependency}: This smell was derived from the recurring pattern where some redundant dependencies are left in DL systems and we noticed five out of 59 projects have modification commits to remove dispensable dependencies. Many DL libraries have repetitive functions, so some practitioners might try similar functions in each library and use the one with the best performance. However, this process adds some unnecessary dependencies to the entire system.

\subsection{Prevalence of Python Smells}
We used PySmell to analyze the Python code smells in our selected commits and identified eight Python code smells that were investigated by Hadhemi et al.~\cite{jebnounscent} (shown in Figure~\ref{fig:exsiting_smells}). The most frequently fixed  Python code smells in our dataset are LM, LPL, and CCC and their respective fixing commit percentages are 6.56\%, 6.32\%, and 6.09\%. LTCE and LLF code smell fixing occupy 3.98\% and 3.04\% percentage of all selected commits. And the fixing commit percentages for MNC, LC, and LSC are 1.17\%, 0.94\%, and 0.70\%. %The Average number of commits for five most occurred Python code smell is 5.20\%.

According to Table~\ref{tab:new_smells}, the percentage of selected commits to fix code smell of \textit{Scattered Use of ML Library} and \textit{Unwanted Debugging code} are respectively 13\% and 17\%. And for \textit{Deep God File} and \textit{Jumbled Model Architecture} code smells, the percentage of commits are 9\% and 6\%. The lowest percentage of commits is \textit{Dispensable Dependency} code smell, which is 2\%. By comparing the percentage of commits containing the code smells between our identified code smells and existing Python code smells we see that newly identified code smells are more frequent compared to generic Python code smells.

%The Average number of commits for the five newly identified code smells is 9.40\%. Thus, comparing to Python code smells, our newly identified code smells are more frequently happens in DL systems. 

\observation{Newly identified code smells occur more frequently in our sample than generic Python code smells.}

\subsection{Code Smells Validation}

\textbf{RQ 3: How do practitioners perceive the identified code smells in DL systems?}

\begin{figure}[h] 
    \centering
  \subfloat[Results of code smells frequency of occurrence\label{1a}]{%
       \includegraphics[width=0.95\columnwidth]{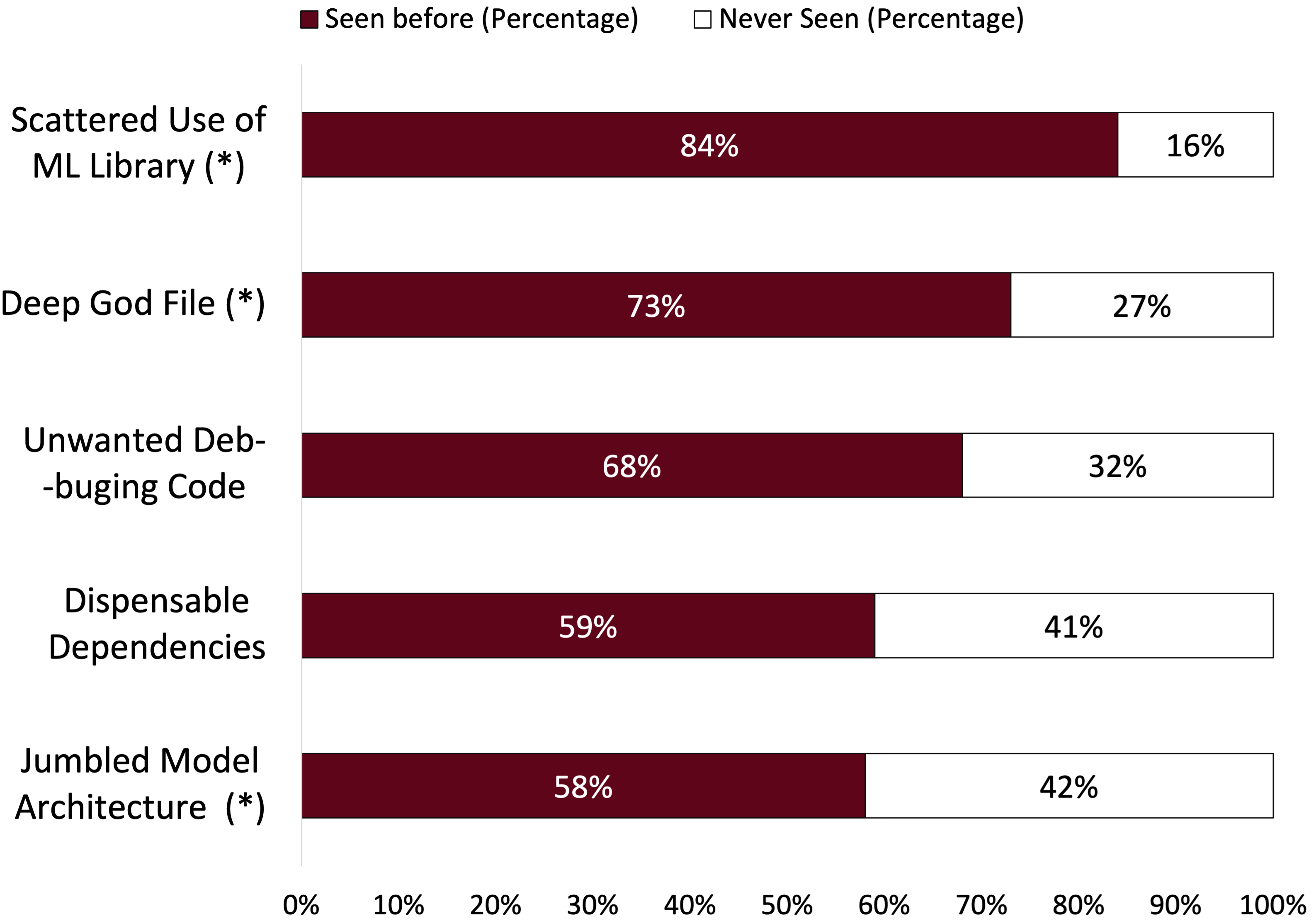}}
    \hfill
    
  \subfloat[Results of code smells impact on deep learning systems\label{1b}]{%
        \includegraphics[width=0.95\columnwidth]{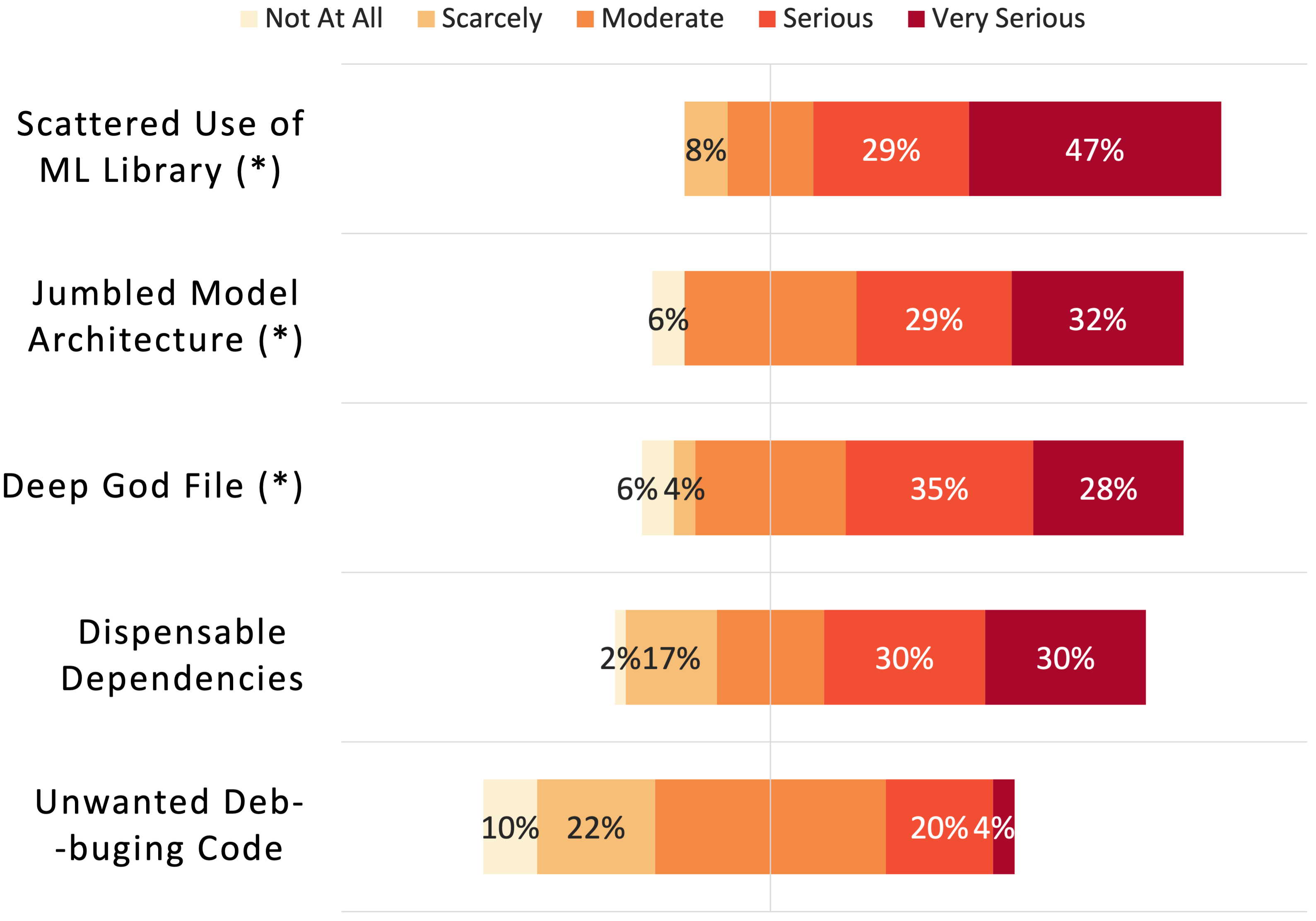}}
  \caption{Aggregated survey results form all respondents}
  \label{fig:codeSmellsValidation} 
\end{figure}

To answer this question, we analyzed the survey results. We asked respondents to what extent they have encountered these code smells and their perception about the impact of these code smells on making DL systems difficult to maintain. The aggregated results are shown in Fig.~\ref{fig:codeSmellsValidation}.

% \boldification{Fig 5 shows the aggregated results for code smells prevalence and damage to maintainability.}

According to the aggregated results shown in Figure.~\ref{fig:codeSmellsValidation}-(a), respondents are familiar with the code smells we identified. 84\% of the respondents expressed that they have seen \textit{Scattered Use of ML Library} code smell before, which matches our repositories mining result that the most frequently occurred code smell is \textit{Scattered Use of ML Library}. The respondents were also familiar with the other code smells. The ranking through mining was closely matched with the survey's ranking as \textit{Unwanted Debugging Code}, and \textit{Deep God File} were among the top three code smells in both rankings.

According to the combined result from all the participants in Figure.~\ref{fig:codeSmellsValidation}-(b), the most impactful code smell is \emph{Scattered Use of ML Library}. More than 60\% of survey respondents reported that these code smells seriously impact their DL systems' maintenance. According to the developers, the other two most impactful code smells are \emph{Jumbled model architecture} and \emph{Deep God File}. Among them, \emph{Deep God File} is also the second most frequent code smell (Figure.~\ref{fig:codeSmellsValidation}).

\observation{According to both mining and developers' responses, \emph{Scattered Use of ML Library} and \emph{Deep God File} are two of the most frequent and impactful code smells.}

\begin{figure}[h]
  \centering
    \includegraphics[width=0.95\columnwidth]{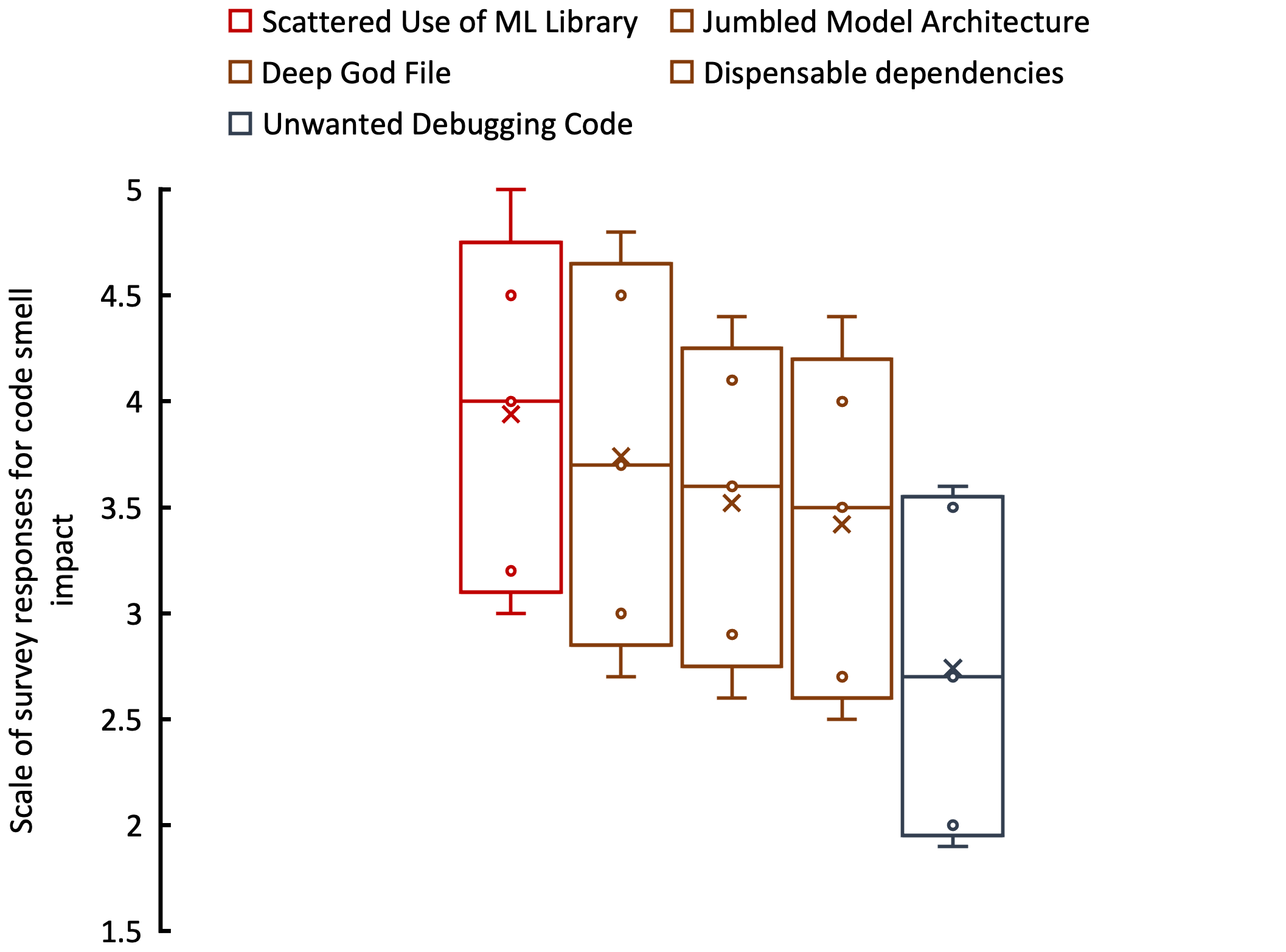}
  \caption{Scott-Knott effect size difference (ESD) test results}
  \label{fig:skESD}
\end{figure}

\begin{figure}[h] 
    \centering
  \subfloat[Survey responses for different roles about code smell Occurrence (shows percentage of respondents have seen code smell before)\label{2a}]{%
       \includegraphics[width=0.95\columnwidth]{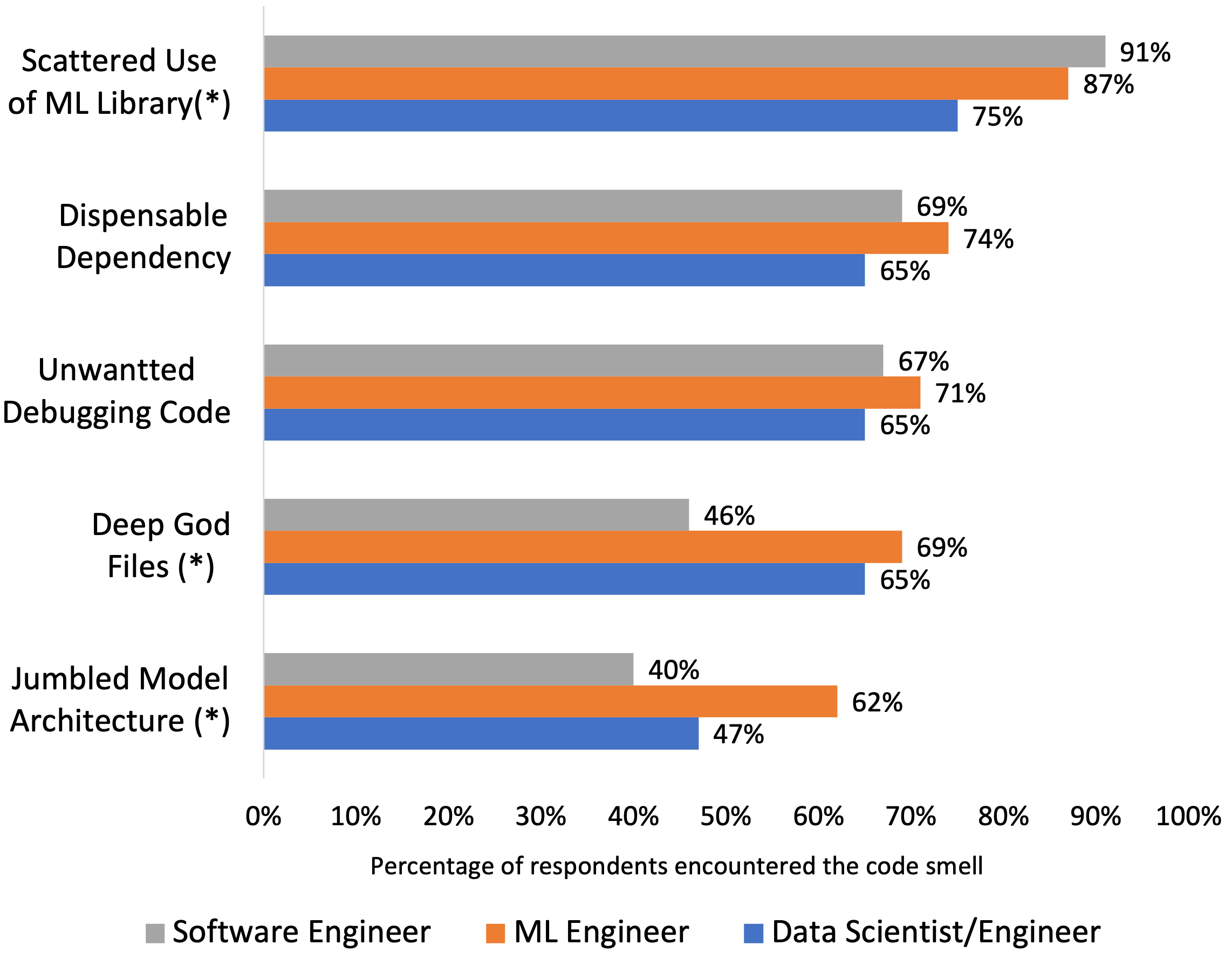}}
    \hfill
    
  \subfloat[Survey responses for different roles about code smell impact (only show the percentage of respondents think the code smell has serious or very serious impact)\label{2b}]{%
        \includegraphics[width=0.95\columnwidth]{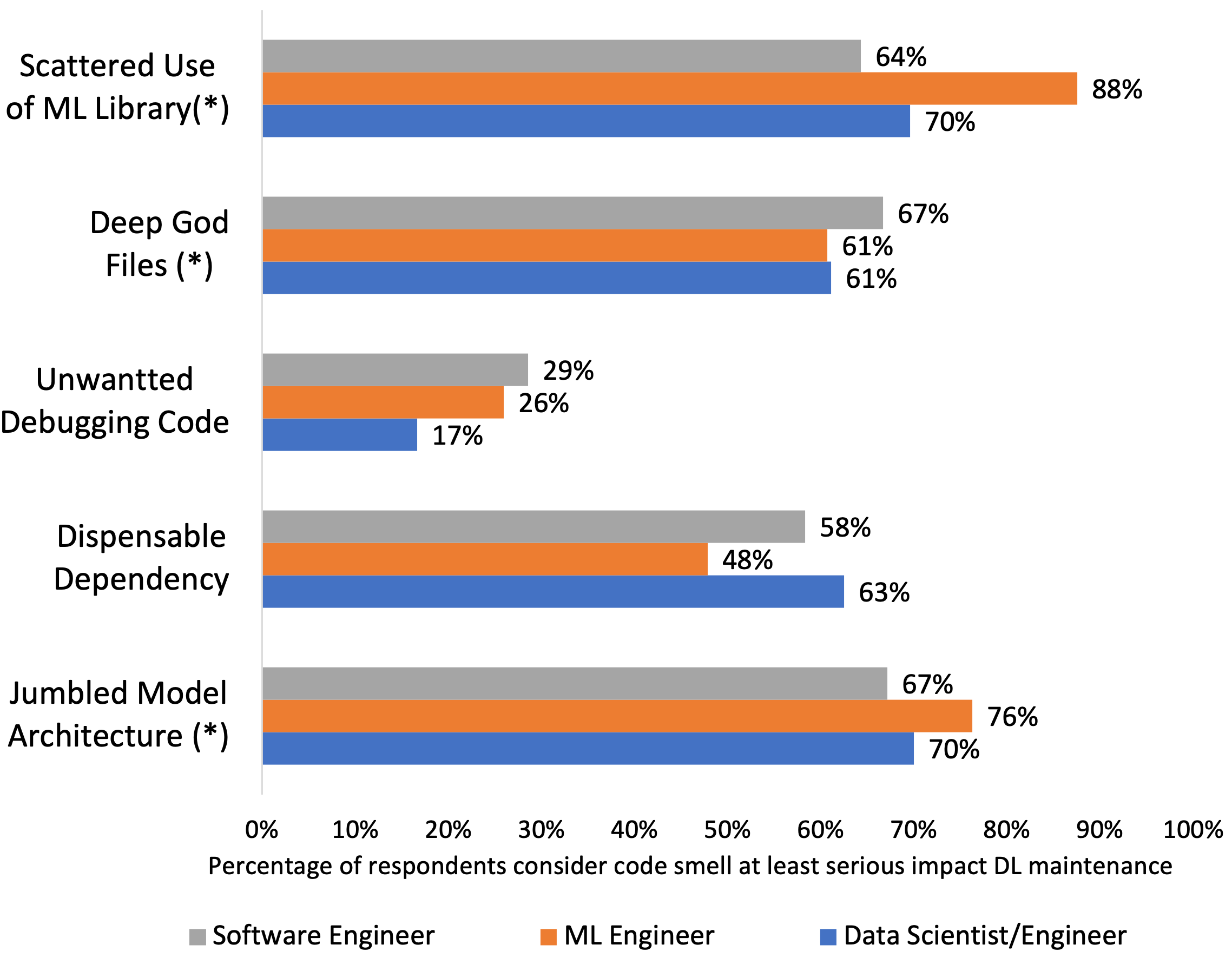}}
  \caption{Separated survey results for different roles}
  \label{fig:codeSmellsSplit} 
\end{figure}

When we grouped the perceived frequency of code smells based on respondents' roles shown in Fig.~\ref{fig:codeSmellsSplit}-(a), the most common code smells for ML engineers, software engineer and data scientist/engineer respondents were \emph{Scattered Use of ML Library, Dispensable Dependency, Unwanted Debugging Code}. However, software engineer met \textit{Scattered Use of ML Library} more often, but ML engineer encountered \textit{Dispensable Dependency} and \textit{Unwanted Debugging Code} more often. It is also reasonable that software engineers and Data engineers encountered less \textit{Jumbled Model Architecture} since they are not primarily maintaining models, but 62\% ML engineer respondents encountered \textit{Jumbled Model Architecture} code smell as they are primarily working with models.

We looked into the impact of these code smells for each role, shown in Fig.~\ref{fig:codeSmellsSplit}-(b), which only shows the percentage of respondents who identified the code smell having serious or very serious impact on system maintenance. The \textit{Scattered Use of ML Library} code smell is considered as the most severe by all three roles, especially by ML Engineers since 88\% of ML Engineer respondents think this code smell has a ``serious impact'' on their system maintenance. Similarly, \textit{Jumbled Model Architecture} is considered as a severe code smell by all three roles, even though it's not common for software and data scientist/engineer. In our analysis, we found that \textit{Unwanted Debugging Code} is a common code smell, but most of respondents do not think it is a severe issue. 

\observation{Different roles encounter code smells differently and they also have varied opinions about the impact of the code smell.}

We conducted Scott-Knott ESD test on the responses collected for pertaining to the impact of code smells on DL maintenance to check if there is significant difference among all newly identified code smells. Figure~\ref{fig:skESD} shows that Scott-Knott ESD categorized five code smells into three different groups. \textit{Scattered Use of ML Library} is categorized in the first group as the most impactful code smell; \textit{Jumbled Model Architecture, Deep God File, and Dispensable Dependency} code smells are categorized into the second group; and \textit{Unwanted debugging Code} is categorized into the third group.

%!TEX root = main.tex

\section{Discussion and Implications}
\label{sec:discussion}

In this section, we discuss the results presented in the previous section and present mitigation strategies, probable root causes for code smells, and practical implications of our study for researchers, educators, tool builders, and developers.

\subsection{Mitigation Strategies}

\textbf{Scattered Use of ML Library}: DL practitioners should import module as an alias to shorten the ML API message chain. That way, when an ML API is updated, maintainers no longer need to modify the usage of this API call throughout the whole project. Instead, they only need to change the code in module importing parts.

\textbf{Jumbled model architecture}: We suggest DL practitioners clearly separate the parts of the DL model with different functions, so that the model code is easier to understand and maintain. 

% We also recommend developers take advantage model management tools, such as ModelHub~\cite{miao2017modelhub}, to manage DL models.

% We suggest developers code DL model strictly following the model original designed architecture. So that the structure of the code matches the architecture design of the code smells. 

\textbf{Deep God File}: We recommend developers place each part of code into a proper file, and have clear boundaries, such as placing model architecture, training, testing and validating program in separate files. If there is a Deep God File, DL practitioners can employ extract class or move function refactoring operations to separate components in such files. 

\textbf{Unwanted Debugging Code}: We advocate DL practitioners remove unused debugging code in a timely manner.  

\textbf{Dispensable dependencies}: We encourage practitioners remove unnecessary dependencies in DL systems since it takes a lot of time and effort to resolve dependency and library version conflicts in DL systems. 

%Apart from the above-identified code smells, we also found other repeatedly occurring modifications, such as replacing hardcoded values and data preparation changes. The hardcoded value does not always make DL systems challenge to maintain, but as the number of the hardcoded values increase, developers should pay more attention to it since it may start making DL systems difficult to maintain. Regarding data preparation modifications in DL systems, Moayad et al.~\cite{alshangiti2019developing} showed that data preprocessing is the most challenging part for DL practitioners. However, it is difficult to provide a general solution for various reasons data preparation modifications, such as data source formatting change, feature change, model input formatting change, etc. Thus, we did not categorize data preparation modification into a code smell.
% \boldification{We saw some issues and make DL practitioners to make modifications during maintenance, but we did not categorize them as code smell, such as hard coded values and etc.}% \textbf{Data preparation changes}: As Moayad et al.~\cite{alshangiti2019developing} showed that data preprocessing is one of the most challenging steps for DL practitioners according to mining Stack Overflow results. We also found that there are 8\% changing commits related to data preparation. But it was difficult to summarize a solution for various reasons behind data preparation modifications. 
\subsection{Probable root causes}

% Team (Background, if you can find citation)
ML teams are composed of different roles with overlapping tasks~\cite{amershi2019software}. We posit that code smells might be a product of such overlapping tasks since the overlap in responsibility leads to unclear maintenance responsibility. A general thought is that this problem is not unique to DL systems but applies to regular systems as well. That is absolutely correct. Nevertheless, this problem is exacerbated in DL systems because of the significantly distinct roles of various team members. Along with creating confusion and dissatisfaction, uncertain responsibilities can result in dropped or mishandled source code and catastrophic consequences down the line. Practitioners need to ensure the maintenance task's boundaries for different roles in DL systems.

% Development process
Differences in job responsibilities among team members can be another reason for accumulating code smells over time. For example, ML engineers mainly concentrate on model development rather than software deployment and maintenance. To obtain a better model performance, they may try different ML libraries and add all tried library dependencies into the system at the same time. Even though ML engineers finally end up requiring only a few of the imported libraries, the unused but imported DL libraries and their dependencies remain in the system. Such unnecessary dependencies introduce additional problems to the software engineers who try to build and maintain the DL system. Since the process used at the ML developer’s end is opaque to the software engineers, it becomes difficult, even impossible in certain cases, for software engineers to remove any unused DL library dependencies. As a result, all the unused dependencies are left in the DL system and the quality of the system as a whole suffers~\cite{sculley2015hidden}. Projects in the industry have started investigating ways to overcome these challenges. One approach is hybrid teams that include ML engineers, data scientists, and DevOps engineers~\cite{MLchallenges}. Further work is needed to help DL systems identify and remove the unused dependencies. Improving cross-team communication, reducing the opaqueness in the development process used within the sub-groups along with ensuring documentation are some of the possible steps to mitigate this to some extent.
% Additional Components and complexity
\subsection{Implication}
\textbf{Implications for researchers, tool builders and educators}: 
Our results show that DL systems have a wide variety of code smells. However, when we looked for code smell related work for DL, we found limited studies. We encourage researchers to investigate many more kinds of code smells in DL systems.

Tool builders can focus on making the code smell detection tools seamlessly integrated into the existing DL development pipeline without causing major disruptions. This is important because research shows that if a workflow is disrupted, practitioners tend to stop using the tool~\cite{johnson2013don}.

The large variety of code smells in DL systems is also good news for educators. Educators can illustrate many design principles by showing both well-designed programs and those that exhibit code smells. Using DL systems as subject case studies is guaranteed to provide a variety of code smells. Moreover, students might also prefer examples from the DL domain given the rise and allure of DL programming.

\textbf{Implications for DL developers}: As Table~\ref{tab:new_smells} shows, identified code smells are distributed in a big percentage of DL systems. Thus, it is important that developers educate themselves about the kinds of code smells that occur in DL systems, and how to mitigate them. Or even better, being conscious about code smells when programming in the first place and avoid them altogether.
%!TEX root = main.tex
\section{Threats to Validity}
\label{sec:threats}
Our refactoring pattern mining was performed on 59 projects carefully selected by Hadhemi et al.~\cite{jebnounscent}. However, these are open source projects, which means the results may not be generalizable to all DL projects, particularly closed-source projects. Nonetheless, the majority of the DL projects use Python, so we believe our code mining on these Python projects still provides significant insights on code smells in DL systems. This is our first step towards building an empirical body of knowledge. With further replication across different contexts by different research teams, we can build a body of knowledge to generalize the results.

The manual analysis applied throughout the study could have introduced unintentional bias. First, we manually identified the commits that were related to maintenance activities based on commit messages and comparing the code before and after an update. Another manual analysis was conducted while grouping the frequently occurring change categories into code smells. This could have introduced bias or mistakes due to the lack of domain expertise. To address this concern, two researchers individually labeled a significant portion of the data. We established a high inter-rater agreement of $0.61$ and $0.83$ respectively for the two manual analyses, which according to Landis et al.\cite{landis_1975}, is considered as a substantial level of agreement and we believe we have minimized this threat.

We ran PythonChangeMiner to obtain frequently changed patterns, and then we used GitcProc to exclude bug fixes. Relying on these tools can be a threat to validity. However, these tools or variant of them has been validated in other studies. We also performed a manual investigation of any refactored code that has not been labeled by GitcProc to identify if there is any systemic error. Through our manual analysis, we did not see any evidence of the systemic error. 

There is a possibility that our participants misunderstood the survey questions. To mitigate this threat, we conducted a pilot study with 11 developers with different background experiences and updated the survey based on the feedback. In order to clarify any confusion, we provided definitions for each of the smells. Additionally, we translated the original survey to simplified Chinese to help native Chinese readers to reduce any confusion. Our survey's language selection and translation process may be subject to bias. It might cause the group of respondents who can read Chinese and English to be over represented. However, it is important to mention that we chose to present our survey in English and Chinese because these are the top two most used languages in software development. Our survey could also have translation errors that cause the questions to deviate from the original meaning. To mitigate these risks, two of the authors (one of them is a native English speaker and the other a native Chinese speaker) discussed the survey and performed the translation together.
%!TEX root = main.tex
\section{Conclusions and future work}
\label{sec:conclusion}
We investigated frequently occurring modifications in DL open source software repositories and identified nine modifications along with five code smells in this work. We also validated the code smells with DL practitioners through a survey. Participants identified the most impactful smells; however, surprisingly, the most frequent code smells are not necessarily the most impactful ones.

Our findings also open up new directions for future research. In addition to the future directions already presented in the discussion and implication sections, future research entails exploring the evolution of the identified code smells and their effect on DL systems' overall quality.

%ractitioners can increase their awareness regarding the presence of these smells and actively take steps to fix them and, whenever possible, prevent them from occurring in the first place, which in the long run will help to ensure the quality of their systems. 

\bibliographystyle{IEEEtranS}
\bibliography{icse}

\end{document}